\documentclass[pra,showpacs,superscriptaddress,twocolumn,10pt]{revtex4-1}
\usepackage[utf8]{inputenc}
\usepackage{braket}
\usepackage[cm]{fullpage}
\usepackage{graphicx}	
\usepackage[english]{babel}
\usepackage{amsmath}
\usepackage{amssymb}
\usepackage{cancel}
\usepackage{natbib}
\usepackage{comment}
\usepackage{color}
\usepackage[normalem]{ulem}
\usepackage{bbold}
\usepackage[outdir=./]{epstopdf}
\usepackage{here}

\DeclareMathOperator{\tr}{tr}
\DeclareMathOperator{\rk}{rk}
\DeclareMathOperator{\im}{Im}

\def\calH{\mathcal{H}}

\def\cL{\mathcal{L}}
\def\bx{x}
\def\idmat{\mathbb{1}}
\def\pt{\textrm{PT}}
\def\pl{\textrm{pl}}
\def\eij{\ket{i}\bra{j}}
\def\nbpart{p}

\makeatletter
	\renewcommand{\maketag@@@}[1]{\hbox{\m@th\normalsize\normalfont#1}}
\makeatother
\begin{document}

\author{N.~Milazzo}
\affiliation{Institut f\"ur theoretische Physik, Universit\"at T\"{u}bingen, 72076 T\"ubingen, Germany}
\affiliation{Université Paris-Saclay, CNRS, LPTMS, 91405, Orsay, France}
\author{D.~Braun}
\affiliation{Institut f\"ur theoretische Physik, Universit\"at T\"{u}bingen, 72076 T\"ubingen, Germany}
\author{O.~Giraud}
\affiliation{Université Paris-Saclay, CNRS, LPTMS, 91405, Orsay, France}

\begin{abstract}
We consider the problem of separability of quantum channels via the Choi matrix representation given by the Choi–Jamiołkowski isomorphism. We explore  three
classes of separability across different cuts between systems and ancillae and we provide a solution based on the mapping of the coordinates of the Choi state (in a fixed basis) to a truncated moment sequence (tms) $y$. This results in an algorithm which gives a separability certificate using semidefinite programming. The computational complexity and the performance of it depend on the number of variables $n$ in the tms and on the size of the 
moment matrix $M_t(y)$ of order $t$.
We exploit the algorithm to numerically investigate separability of families of 2-qubit and single-qutrit channels; in the latter case we can provide an answer for examples explored earlier through the criterion based on the negativity $N$, a criterion which remains inconclusive for Choi matrices with $N=0$.
\end{abstract}


\title{Truncated moment sequences and a solution to the channel separability problem}

\date{\today}

\maketitle

\section{Introduction}

Describing entanglement properties of quantum states has been at the center of many investigations in the recent years. In that context, it is of high relevance to understand the way entanglement evolves under physical operations acting on quantum states \cite{Intro1, Intro2, Intro3, Intro4, Intro5, CirDurKra01}. 
The mathematical object associated with a physical operation is a quantum channel, which acts on the joint state of a system $\mathcal{A}$ and its environment to produce an output state. The environment can be seen as an ancilla system $\mathcal{A'}$, with which the system  $\mathcal{A}$ is possibly entangled. The system  $\mathcal{A}$ itself may be bipartite and made of two subsystems $A$ and $B$ which may or may not be entangled with one another, or with their respective ancillae 
$A'$ and $B'$. Since a channel acts on both the system and its ancilla, the output state may be entangled in different ways, which leads to different definitions of separability
of quantum channels 
\cite{NJPhD, EB, FilZim13, RahJaqPau18, Christandl19}. These definitions depend on whether the total state of the system and ancilla is separable for instance across the cut $\mathcal{A}-\mathcal{A'}$, or across the cut $A-B$.

The Choi-Jamio\l{}kowsi isomorphism relates completely positive trace-preserving maps with density matrices, or equivalently completely positive maps with positive operators. Characterizing separability for channels can be investigated in the light of results obtained for quantum states. Many theoretical results have been obtained for states in terms of separability criteria. One of the most well-known necessary conditions for separability is the PPT criterion, which states that if a state $\rho$ is separable then $\rho^\pt\geq 0$, with $\rho^\pt$ the partial transpose with respect to one of the subsystems \cite{peres96,Horodecki96}.

 As was shown recently \cite{tms17}, the separability problem for states can be recast as a "truncated moment" problem, a problem well-studied in recent years in the mathematical literature. The truncated moment problem consists in finding conditions under which a given sequence of numbers corresponds to moments of a probability distribution. The moment problem corresponds to the case where an infinite sequence is given, while in the truncated moment problem only the lowest moments are fixed and the aim is to find a measure matching these moments. Of relevance for the separability problem, as we will see, is the $K$-truncated moment problem, where the measure is additionally required to have the set $K$ as support. In \cite{tms17} we showed that asking whether a quantum state is separable along an arbitrary partition of Hilbert space can be cast in the form of a  $K$-truncated moment problem, and we applied this approach to symmetric multiqubit states. 

In the present paper our goal is to apply this formalism to the more general situation of the separability of quantum channels. 
We provide theorems that give necessary and sufficient conditions for a channel to be separable or entanglement breaking, as well as an algorithm that implements the theorems numerically. 
We then consider various examples of detection of separability in quantum channels.

\section{Definitions}
We start by recalling some elementary definitions.

\subsection{Quantum channels}

 Let $ \rho $ be a quantum state acting on a tensor product $ H=H^{(1)}\otimes ... \otimes H^{(d)} $ of Hilbert spaces $ H^{(i)}$ of finite dimension.
 Any physical transformation can be described by a completely positive map, that is, a map $\Phi$ such that $\Phi\otimes\mathbb{1}$ is positive on all states acting on an extended Hilbert space $H\otimes H'$ (where $H'$ is the Hilbert space of an ancillary system of arbitrary size). A quantum channel $\Phi$ is therefore defined as a completely positive trace-preserving linear map, which maps $\rho$ to a state $\rho'=\Phi(\rho)$ acting on some Hilbert space (that for simplicity we consider here equal to $H$, so that $\Phi:\cL(H)\to\cL(H)$, where $\cL(H)$ is the set of linear operators on $H$). 

Let $N$ be the dimension of the Hilbert space $H$. A density matrix can be expanded as $\rho=\sum_{i,j}\rho_{ij}\eij$, with $\ket{i}$ the vectors of the canonical basis of $H$. To any linear map $\Phi$ mapping $\rho$ to $\rho'$ one can associate a superoperator $M$ of size $N^2$ such that $\rho'_{ij}=M_{ij,kl}\rho_{kl}$ (with summation over repeated indices), and a dynamical matrix $D_{\Phi}$ defined  \cite{Dynamat61} by a reshuffling of entries of $M$, namely $(D_\Phi)_{ij,kl}=M_{ik,jl}$ \cite{Zycbook}. Alternatively one can define the Choi matrix 
\begin{equation}
\label{defchoi}
C_{\Phi}=\sum_{i,j} \Phi(\eij)\otimes \eij
\end{equation}  \cite{Choimat75}, which coincides with $D_\Phi$ when written in the canonical basis. 
The Choi matrix $C_{\Phi}$ is Hermitian.
The map $\Phi$ is positive if and only if  the corresponding Choi matrix $C_{\Phi}$ is block-positive (that is, positive on product states in $H\otimes H$) \cite{Jam72}. According to Choi's theorem \cite{Choimat75}, $\Phi$ is completely positive if and only if its Choi matrix is positive semidefinite. Finally, $\Phi$ is trace-preserving if and only if the $N^2$ conditions $\sum_i(C_{\Phi})_{ij,il}=\delta_{jl}$ are fulfilled. These conditions imply that $\tr C_\Phi=N$.

As a consequence, if $\Phi$ is a quantum channel, then $\frac{1}{N}C_{\Phi}$ can be seen as a density matrix acting on $H\otimes H$. Any completely positive trace-preserving map can be associated with a density matrix in that way. The Choi-Jamio\l{}kowsi isomorphism is a bijection between a quantum channel $\Phi$ and its Choi matrix $C_\Phi$ \cite{Jam72, Zycbook}. We shall also make use of the fact that a quantum channel can 
be written in Kraus form as
\begin{equation}
\label{kraus}
\Phi(\rho)=\sum_{l}E_l\rho E_l^{\dagger}, \qquad \sum_{l}E_l^\dagger E_l=\mathbb{1}\,. 
\end{equation}
The Kraus operators $E_l$ are not unique, but a canonical form can be found by diagonalizing the Choi matrix and reshuffling its eigenvectors into square matrices, in which case a set of at most $N^2$ Kraus operators suffices \cite{Zycbook}.

\subsection{Separability of channels}
\label{defsep}

A bipartite quantum state $\rho$ acting on a Hilbert space $H_{A}\otimes H_{B} $ is separable if it admits a decomposition
\begin{equation}
\rho=\sum_iw_i\rho^{(A)}_i\otimes\rho^{(B)}_i
\end{equation}
with $w_i\geq 0$ and $\rho^{(A)}_i,\rho^{(B)}_i$ acting on $H_{A}, H_{B} $ respectively. More generally, a 
positive semidefinite
matrix $M$ is said to be separable if it can be written as 
\begin{equation}
\label{sepmat}
M=\sum_k P_k\otimes Q_k
\end{equation}
with $P_k$ and $Q_k$ positive semidefinite matrices.

Various kinds of channel separability have been introduced in the literature. Consider the Hilbert space $H=H_{A}\otimes H_{B} $ describing a system partitioned into two subsystems $A$ and $B$, and let $\Phi: \cL(H_{A}\otimes H_{B})\rightarrow \cL(H_{A}\otimes H_{B})$ be a completely positive map. 
As a criterion for complete positivity one must consider the extended Hilbert state $H\otimes H'$
with $H'=H$, where here and 
in the following the prime is used to denote the ancilla system. The corresponding Choi matrix $C_{\Phi}$ can be seen as a density matrix acting on Hilbert space $\calH=  H_{A}\otimes H_{B}\otimes H_{A'}\otimes H_{B'}$. Following Eq.~\eqref{defchoi} it can be expressed as $C_{\Phi}=\sum_{ijrs}\Phi(\ket{ir}\bra{js})\otimes \ket{ir}\bra{js}$.\\

{\it \textbf{Separable channels.}} 
 $ \Phi$ is called separable if it takes the form $\Phi(\rho)=\sum_l (A_{l}\otimes B_{l})\rho (A_{l}\otimes B_{l})^{\dagger}$ \cite{NJPhD}. In other words, the Kraus operators for the channel $ \Phi $ in \eqref{kraus} can be factored as $E_l=A_l\otimes B_l $. Such channels map separable states to separable states. In terms of these Kraus operators, the Choi matrix of a separable map $ \Phi $ is given by
  \begin{equation}
 C_{\Phi}=\sum_{i,j,r,s} \sum_l  A_l \vert i\rangle\langle j\vert A_l^{\dagger} \otimes  B_l \vert r\rangle\langle s\vert B_l^{\dagger}\otimes \vert i\rangle\langle j\vert \otimes \vert r\rangle\langle s\vert .
 \label{choieb}
 \end{equation}
 Swapping $H_{A'}$ and $H_{B} $  we can interpret  $ C_{\Phi} $ as an operator in $H=  H_{A}\otimes H_{A'}\otimes H_{B}\otimes H_{B'} $ and reexpress it as 
 \begin{equation}
 C_{\Phi}=\sum_l\sum_{i,j} A_l \vert i\rangle\langle j\vert A_l^{\dagger}\otimes \vert i\rangle\langle j\vert   \otimes \sum_{r,s} B_l \vert r\rangle\langle s\vert B_l^{\dagger}\otimes \vert r\rangle\langle s\vert.
 \label{choisep}
 \end{equation}
 It is clear that $ \sum_{i,j}  A_l \vert i\rangle\langle j\vert A_l^{\dagger}\otimes \vert i\rangle\langle j\vert $ is positive semidefinite for all $ l $, because it is the Choi matrix of the completely positive map $\rho \mapsto  A_l \rho A_l^{\dagger}$; and the same holds for $B$. Therefore, $C_{\Phi}$ can be written as a sum $\sum_l M_\mathcal{A}^{(l)}\otimes M_\mathcal{B}^{(l)}$ with $M_\mathcal{A}^{(l)}$ and $M_\mathcal{B}^{(l)}$ positive semidefinite: it is thus a separable matrix across the $(A-A')-(B-B')$ cut. It was shown in \cite{CirDurKra01} that the converse is true, namely $ C_{\Phi} $ is separable across the $(A-A')-(B-B')$ cut if and only if $ \Phi $ is a separable map. We shall use this characterization of separable channels in Section \ref{thm}.
 
We will call $\Phi$  {\bf fully separable} (FS) if the corresponding $ C_{\Phi} $ is separable across all possible cuts.\\

\begin{figure}
\includegraphics[scale=0.33]{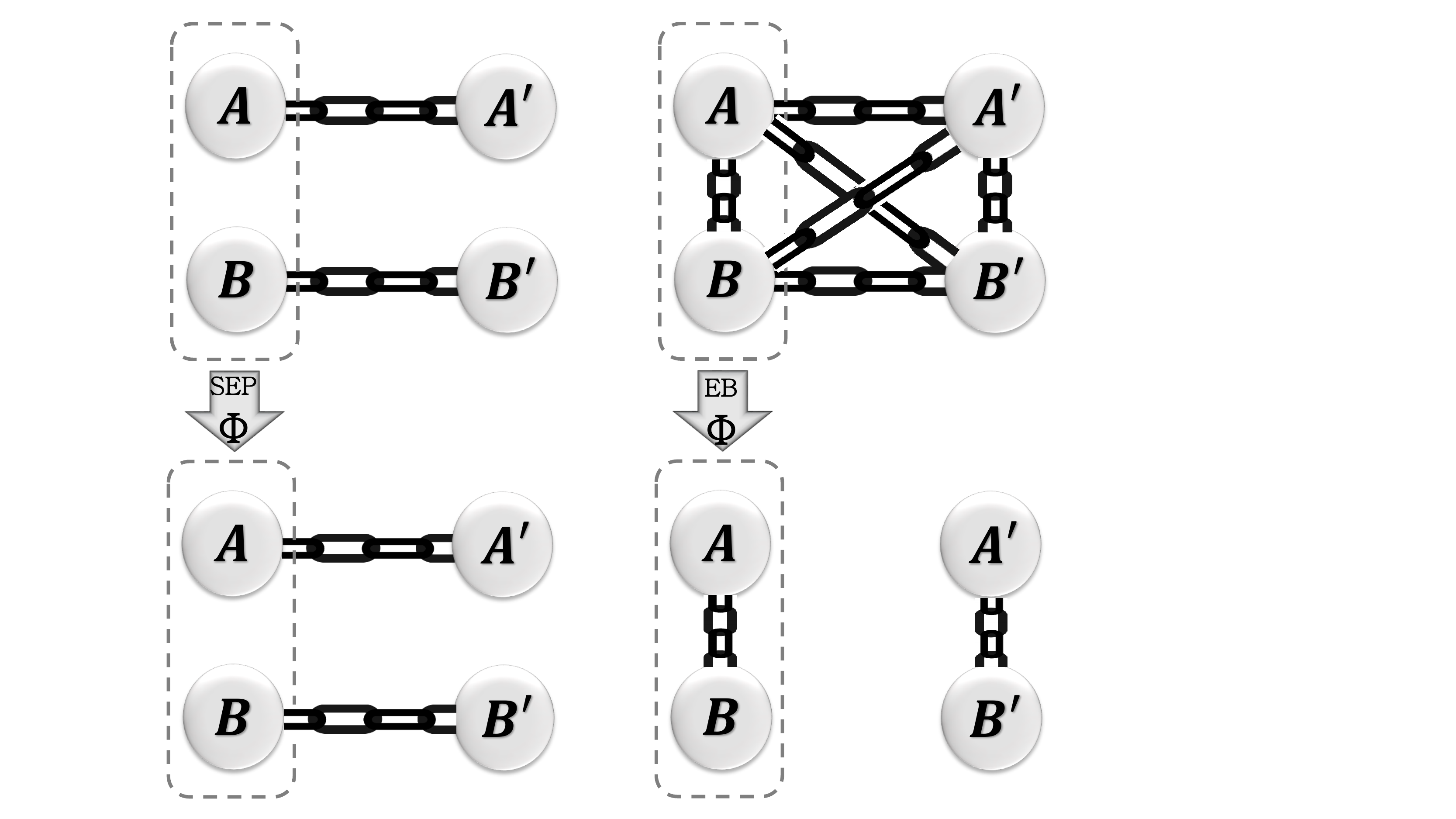}
\caption{Difference between separable (left) and entanglement-breaking (right) channels for a bipartite system $AB$ with ancillae $A'B'$. The chains represent entanglement. A separable channel preserves separability between  ($ A-A' $) and ($ B-B' $), while an entanglement breaking channel destroys entanglement between $ A $ and all the ancillae and $ B $ and all the ancillae, giving separability between ($ A-B $) and ($ A'-B' $).}
\label{one}
\end{figure}

{\it \textbf{Entanglement-breaking channels.}} $\Phi$ is called entanglement breaking (EB) 
\cite{EB} if $ (\Phi\otimes \idmat)(\rho) $ is a separable state across the $H-H'$ cut whatever the initial state $\rho\in\cL(\calH)$. It does not address the separability of the bipartite system $H$ into $A$ and $B$, but rather the separability between the system and its environment (it can therefore be defined for one-qubit channels). Various necessary and sufficient conditions for entanglement breaking have been obtained in \cite{EB}. One necessary and sufficient criterion is that there exist a Kraus form where all Kraus operators have rank 1. In terms of the Choi matrix, a necessary and sufficient condition
for EB is that $ C_{\Phi} $ be separable across the $(A-B)-(A'-B')$ cut. Physically these channels correspond to the case in which the output state is prepared according to the measurement outcomes made by the sender and sent via a classical channel to the receiver. We point out the difference between separable and entanglement-breaking channels in Fig.~\ref{one}.

Channels which become entanglement breaking after a sufficient number of compositions
with themselves are called  \textbf{eventually entanglement breaking channels} \cite{RahJaqPau18, Christandl19}.\\

{\it \textbf{Entanglement annihilating channels.}} $\Phi$ is called entanglement annihilating \cite{EA1} if it destroys any entanglement within the system $H$ (but it does 
not necessarily destroy entanglement between $H$ and $H'$). A necessary and sufficient condition
for entanglement annihilating channels 
in terms of the Choi matrix is that $C_{\Phi}\geq 0$ and that its partial trace over $A$ and $B$ is proportional to the identity matrix (see Corollary 1 of \cite{FilZim13}). Such a condition on partial trace is not implementable in tms form, so we will not address this type of separability.\\

\section{Truncated moment sequences}
\label{sectms}

\subsection{The tms problem}
In order to be as self-contained and pedagogical as possible for a physics-oriented audience, we start by reviewing and explaining some results from the mathematical literature \cite{CurFia96, CurFia00, CurFia05, LaurentReview2010, HeltonNie2012, Nie14, NieZhang16}.
We follow the nice presentation from \cite{Lau05}. We then recall the theorems obtained in \cite{tms17} for quantum states, and formulate them in the case of quantum channels.

A truncated moment sequence (tms) $y=(y_{\alpha})_{\vert\alpha\vert \leq 2d}$ of degree $2d$ is a finite set of real numbers indexed by $ n $-tuples $ \alpha=(\alpha_1, ..., \alpha_n) $ of integers $ \alpha_i\geq 0$ such that $ \vert \alpha\vert=\sum_i \alpha_i \leq 2d$ (here we only consider tms of even degree: indeed, although the definition would extend trivially to odd-degree tms, even-degree tms are the only ones involved in the theorems below, so this slightly simplifies notations). We denote by $S_{2d}$ the set of $n$-tuples $\alpha=(\alpha_1, ..., \alpha_n)$ with $ \vert \alpha\vert\leq 2d$, so that $y$ is a vector in $\mathbb{R}^{S_{2d}}$. The number of such $n$-tuples is
\begin{equation}
\label{nrnuples}
\sum_{k=0}^{2d} \binom{k+n-1}{n-1}=\binom{n+2d}{2d}.
\end{equation}
A moment sequence corresponds to a situation where all $y_{\alpha}$ are known to arbitrary order, which we denote by $y\in \mathbb{R}^{S_{\infty}}$.

The truncated moment problem (tms problem) is the problem of finding whether there exists a representing measure for a given sequence $y$, that is, a positive measure $d\mu$ such that 
$y_\alpha=\int x^\alpha d\mu(x)$ for all $\alpha$ with $ \vert \alpha\vert\leq 2d $. Here the notation $ x^{\alpha}$ stands for $\prod_{i=1}^{n}x_i^{\alpha_i}$. 

The $K$-tms problem addresses the case where the measure $d\mu$ is additionally required to be supported by 
a semialgebraic set $K$, that is, a set defined by polynomial inequalities.
We shall use the notation $ K=\lbrace \bx \in \mathbb{R}^n\vert g_1(\bx)\geq 0, ...,  g_m(\bx)\geq 0\rbrace$ with $  g_j(\bx) $ multivariate polynomials. The sequence $y$ has a representing measure for the $K$-tms problem if for all $\alpha$ with $|\alpha|\leq 2d$
\begin{equation}
\label{ktms}
 y_\alpha =\int_K x^{\alpha} d\mu(\bx).
\end{equation}
Necessary and sufficient conditions for the solution of the tms problem can be obtained in terms of moment matrices. Given a tms $ (y_{\alpha})_{\vert\alpha\vert \leq 2d}$, its moment matrix of order $t$ is the matrix $M_t(y)$ indexed by $\alpha,\beta$ with $\vert\alpha\vert, \vert\beta\vert \leq t$ and defined 
as $M_t(y)_{\alpha\beta}=y_{\alpha+\beta}$. The entries of the matrix involve indices of $y$ up to order $2t$, and since the highest index of $y$ is $2d$  (by definition of the tms) 
such a matrix is defined only if $t\leq d$. The size of $M_t(y)$ is given by the number of moments up to order $t$, that is, $\binom{n+t}{t}$. In the case of an infinite moment sequence, the matrix $M(y)$ is infinite. 

Necessary and sufficient conditions for the 
solution of the 
$K$-tms problem additionally involve the localizing matrices associated with polynomials $g_j$ specifying $K$, which are defined as follows. Any polynomial $g$ of $n$ variables $x_1,...,x_n$ 
can be decomposed over monomials as $g=\sum_{|\alpha|\leq \deg(g)} g_\alpha \bx^\alpha$. It can thus be seen as a vector in $\mathbb{R}^{S_{\deg(g)}}$. For a tms $(y_{\alpha})_{\vert\alpha\vert \leq 2d}$ and a polynomial $g$, we define a shifted sequence $g\star y$ by setting $(g\star y)_{\alpha}=\sum_{\gamma} g_{\gamma} y_{\alpha+\gamma}$. The localizing matrix of order $t$ associated with $g$ is defined as the moment matrix of order $t$ of the shifted sequence, that is, $M_t(g\star y)$. Explicitly, its components read $M_t(g\star y)_{\alpha\beta}=\sum_{\gamma} g_{\gamma} y_{\alpha+\beta+\gamma}$. The highest index of $y$ involved here is $2t+\deg(g)$, so that the matrix is defined only for $2t+\deg(g)\leq 2d$, that is, $t\leq d-\deg(g)/2$. The $m$ polynomials defining $K$ give rise to $m$ localizing matrices $M_t(g_j\star y)$. In order that all of them be defined, the order $t$ has to be such that $t\leq d-d_0$
with 
\begin{equation}
\label{d0}
d_0=\max_{1\leq j \leq m}\lbrace 1,\lceil \deg(g_j)/2\rceil\rbrace, \end{equation}
that is, the degree of $y$ has to be greater than 
or equal to 
$2(t+d_0)$.

The three theorems below give necessary and sufficient conditions for a tms (or a full moment sequence) to have a representing measure, supported on $K$ or not. In all cases, the representing measure is $r$-atomic, meaning that it is a sum of $r$ delta functions with positive weights, $d\mu(\bx)=\sum_j \omega_j \delta(\bx-\bx_j)$. The central criterion is the existence of extensions. An extension of a tms $y$ of degree $2d$ is a tms of degree $2d'$ with $d'>d$ whose restriction to indices of order $2d$ or less coincides with $y$. We denote it again by $y$. One can define the moment matrix of order $t$ of such an extension for all $t\leq d'$, and we then say that for $t'>t$, $M_{t'}(y)$ is an extension of $M_{t}(y)$. An extension $M_{t'}(y)$ is said to be a flat extension of $M_{t}(y)$ if it satisfies the condition that its rank is equal to the rank of $M_{t}(y)$, that is,
\begin{equation}
\rk M_{t'}(y)=\rk M_{t}(y).
\label{flat}
\end{equation}
In particular, if \eqref{flat} holds then $M_{t'}(y)\geq 0 \Leftrightarrow M_{t}(y)\geq 0$ (see Appendix \ref{proofrank}).
Theorem 1 below deals with the moment problem, Theorem 2 with the tms problem and Theorem 3 with the $K$-tms problem.\\

{\em Theorem 1.} (\cite{CurFia96}; see theorem 1.2 of \cite{Lau05}) Let $y\in \mathbb{R}^{S_{\infty}}$. If $M(y)\geq 0$ and $\rk M(y)=r$ is finite, then $y$ has a unique representing measure, which is $r$-atomic.\\

{\em Theorem 2.} (\cite{CurFia96}; see theorem 1.3 and Corollary 1.4 of \cite{Lau05}) Let $y\in \mathbb{R}^{S_{2t}}$. If $M_t(y)\geq 0$ and $M_t(y)$ is a flat extension of $M_{t-1}(y)$, then $y$ can be extended to $y\in \mathbb{R}^{S_{2t+2}}$ in such a way that $M_{t+1}(y)$ is a flat extension of $M_t(y)$. \\

From induction and using Theorem 1, one concludes that the tms in $\mathbb{R}^{S_{2t}}$ can be in fact extended to $y\in \mathbb{R}^{S_{\infty}}$ and has a unique representing measure, which is $r$-atomic with $r=\rk M_t(y)$. Moreover one can show (see \cite{Lau05} for detail) that the $r$ atoms $\bx_i$ which support the measure can be obtained from the kernel of $M_t(y)$, that is, the set of polynomials $p=\sum_\alpha p_\alpha\bx^\alpha$ such that $\sum_\beta M_t(y)_{\alpha\beta}p_\beta=0$. More specifically, the set of $\bx_i$ is the variety $\mathcal{V}(\ker M_t(y))=\{x\in\mathbb{C}^n; f(x)=0\ \forall\ f\in\ker M_t(y)\}$, that is, the set of common roots of polynomials in the kernel of $M_t(y)$. In words, what the above results say is that in order to find a representing measure for $y\in\mathbb{R}^{S_{2d}}$ one has to start from the moment matrix $M_{t=d}(y)$ (which is the smallest moment matrix containing all the data) and look for extensions of higher and higher order, until for some order $t$ one has $\rk M_t(y)=\rk M_{t-1}(y)$. If such an extension exists then the representing measure exists and is supported by the common roots of polynomials of $\ker M_t(y)$.\\

{\em Theorem 3.} (\cite{CurFia96}; see theorem 1.6 of \cite{Lau05})  Let $y\in \mathbb{R}^{S_{2t}}$ and $r=\rk M_t(y)$. Then $y$ has a  $r$-atomic representing measure supported on $K$
if and only if $M_t(y)\geq 0$ and there exists a flat extension $M_{t+d_0}(y)$ with $M_t(g_j\star y)\geq 0$ for $1\leq j \leq m$, and $d_0$ defined in \eqref{d0}.

This theorem can be decrypted as follows. Starting from the moment matrix of order $d$ and looking for higher-order extensions of order $t$, if there exists an extension $M_{t+d_0}(y)$ with $\rk M_{t+d_0}(y)=\rk M_{t}(y)=r$ then all its submatrices $M_{t+1}(y), M_{t+2}(y),...$ are also flat extensions of $M_t(y)$. From theorems 1 and 2 one readily concludes that there exists a unique $r$-atomic representing measure; the atoms are given by the variety associated with the kernel of the first extension where the flatness condition is achieved. However these atoms may not be located on $K$. The conditions $M_t(g_j\star y)\geq 0$ on the localizing matrices precisely enforce that additional condition (see Appendix \ref{proofth3} for an insight into the proof). As mentioned above, these matrices are only defined if the degree of $y$ is greater than $2(t+d_0)$, which is why, in order to fulfill these conditions, one has to find extensions in $y\in \mathbb{R}^{S_{2(t+d_0)}}$. Therefore, although an extension to $M_{t+1}(y)$ is enough to guarantee the existence of a $r$-atomic representing measure, an extension to $M_{t+d_0}(y)$ is required so that it is supported by $K$. As a consequence, achieving the flatness condition requires to go quickly to matrices of high order, which has an impact in terms of computational complexity.\\

\subsection{Tms for quantum states}
\label{tmsStates}
Let us now apply these theorems to quantum states, following \cite{tms17}.
Consider a quantum state $\rho$ acting on the tensor product $ H=H^{(1)}\otimes ... \otimes H^{(\nbpart)} $ of Hilbert spaces $H^{(i)}$ with $\dim \mathcal{L}(H^{(i)})=\kappa_i+1$. Let $ S^{(i)}_{\mu_{i}} $ ($ 0\leq \mu_i \leq \kappa_i $) be a set of Hermitian matrices forming an orthogonal basis for $\mathcal{L}(H^{(i)})$, and  $S_{\mu_{1}\mu_{2}...\mu_{\nbpart}} =S^{(1)}_{\mu_{1}}\otimes ...\otimes S^{(\nbpart)}_{\mu_{\nbpart}}$  an orthogonal basis of $\mathcal{L}(H)$. We expand $\rho$ as
\begin{equation}
 \rho=  X_{\mu_{1}\mu_{2}...\mu_{\nbpart}}   S_{\mu_{1}\mu_{2}...\mu_{\nbpart}}
 \label{state}
\end{equation}
(with implicit summation over repeated indices), where $X_{\mu_{1}\mu_{2}...\mu_{\nbpart}}=\tr(\rho S_{\mu_{1}\mu_{2}...\mu_{\nbpart}})$ are the (real) coordinates of the state. Here each index $\mu_i$ runs from 0 to $\kappa_i$, and we will use latin letters $a_i$ for indices running from 1 to $\kappa_i$. It will prove convenient to take $S^{(i)}_0$ as the identity matrix of size the dimension of $H^{(i)}$. 
Actually, as detailed in \cite{tms17}, the matrices $S_{\mu_{1}\mu_{2}...\mu_{\nbpart}}$ need not be an orthogonal basis: it suffices that they be a tight frame (a mathematical structure bearing some analogy with orthogonal bases), which proves useful for example in the case of symmetric states, where some redundancy of the matrices in the expansion \eqref{state} is handy.

One can associate with $\rho$ a tms $ y=(y_{\alpha})_{\vert\alpha\vert \leq \nbpart} $ of degree $ \nbpart $ in the following way.  A density matrix acting on Hilbert space $H^{(i)}$ can be expanded as $\sum_{\mu_i=0}^{\kappa_i} x_{\mu_i}^{(i)}S^{(i)}_{\mu_{i}}$. We associate to $H^{(i)}$ a set of $\kappa_i$ variables $x_{a_i}^{(i)}$, $1\leq a_i\leq \kappa_i$. Let $x=(x_1, x_2,...,x_n)$ 
be the vector of all these variables. In the general case $(x_1, x_2,...,x_n):=(x_{1}^{(1)},x_{2}^{(1)},\ldots, x_{\kappa_p}^{(p)})$ and $n=\sum_i\kappa_i$, and each $x_k$ corresponds to a certain $x_{a_i}^{(i)}$, whereas if we consider symmetric states (i.e.~mixtures of pure states invariant under permutation of the $H^{(i)}$) only one set of variables, say $x_{a_1}^{(1)}$, should be considered, and then $n$ is the common value $\kappa_1=\kappa_2=\ldots$.

An arbitrary monomial of these variables $x_k$ can be written as $x^{\alpha}\equiv\prod_{k=1}^{n}x_k^{\alpha_k}$, where $\alpha_k$ counts the number of variables $x_k$ in the monomial. We then define a tms by $ y_\alpha =X_{\mu_{1}\mu_{2}...\mu_{\nbpart}}$, where $\alpha$ is the index such that 
$x^\alpha=\prod_{i=1}^p x_{\mu_i}^{(i)}$.
Since $X$ has $\nbpart$ indices we have $|\alpha|\leq \nbpart$, 
so that $y_\alpha$ is a tms of degree $\nbpart$. In fact, in order to define a moment matrix, an even-degree tms is required. Thus we set $\nbpart=2d$ if $\nbpart$ is even or $p=2d-1$ if $\nbpart$ is odd. Thus, $ X_{\mu_{1}\mu_{2}...\mu_{\nbpart}} $ is mapped to a tms $ (y_{\alpha})_{\alpha \leq 2d} $ (and in the case where $\nbpart$ is odd the moments of order exactly $2d$ remain unspecified).

As an example, let us consider the case of a 
 state of two spins-1.
We expand it as $ \rho=  X_{\mu_{1}\mu_{2}}  S_{\mu_{1}\mu_{2}}$, where indices $\mu_i$ run from 0 to 8 (since a spin-1 density matrix is a $3\times 3$ Hermitian matrix and can be described by 9 real numbers). We then introduce the vector of variables $x=(x_1,x_2,...,x_{16})$, where $x_1,...,x_8$ are associated with the first spin and $x_9,...,x_{16}$ with the second. Entries $X_{\mu_{1}\mu_{2}}$ define a tms $y_\alpha$ of degree 2 where each $\alpha$ is a vector of integers of length 16 with all entries equal to 0 if $\mu_1=\mu_2=0$, a single nonzero entry $\alpha_{\mu_1}=1$ if $\mu_1\neq 0$ and $\mu_2=0$, a single entry $\alpha_{\mu_2+8}=1$ if $\mu_2\neq 0$ and $\mu_1=0$, and two entries equal to 1 if both $\mu_1$ and $\mu_2$ are nonzero. Each of these $\alpha$ is associated with a monomial, for instance $X_{3\;8}$ corresponds to  $\alpha=(0,0,1,0,0,0,0,0,0,0,0,0,0,0,0,1)$ or to $x_3x_{16}$.

As shown in \cite{tms17}, the problem of finding whether $\rho$ is separable across the multipartition $H^{(1)}\otimes ... \otimes H^{(\nbpart)} $  is equivalent to a $K$-tms problem.
Indeed, projecting the separability condition on the basis $S_{\mu_{1}\mu_{2}...\mu_{\nbpart}}$, coordinates of a separable state can be written as
\begin{equation}
X_{\mu_{1}\mu_{2}...\mu_{\nbpart}}=\int_K x_{\mu_1}^{(1)}x_{\mu_2}^{(2)}...x_{\mu_\nbpart}^{(\nbpart)} d\mu(\bx)
\label{def}
\end{equation}
with $ x_0^{(i)}=1 $, $ \bx=(\bx^{(1)},\bx^{(2)},...,\bx^{(\nbpart)}) \in \mathbb{R}^n$ ($ n=\sum_i \kappa_i $), $ \bx^{(i)}=(x^{(i)}_a)_{1\leq a \leq t_i} \in \mathbb{R}^{\kappa_i} $ and $d\mu(\bx)=\sum_j \omega_j \delta(\bx-z_j)$ a measure supported on a semialgebraic set $K\subset \mathbb{R}^n$ defined by the positivity of the density matrices on each local Hilbert space (that is, the measure is an "atomic" measure, with "atoms" $z_j\in K$). This tms problem is equivalent to asking whether there exists a positive measure $ d\mu $ with support $ K $ for a tms whose moments are the $y_\alpha$ given as explained above by the coordinates $ X_{\mu_{1}\mu_{2}...\mu_{\nbpart}}$ of the state $\rho$. In this language, Eq.~\eqref{def} precisely takes the form \eqref{ktms}. As a consequence, 
separability of $\rho$ can be addressed in the following way:
given a state $\rho$, we can map its coordinates $ X_{\mu_{1}\mu_{2}...\mu_{\nbpart}} $ to a tms $ (y_{\alpha})_{\alpha \leq 2d}$ and look for extensions $(y_{\alpha})_{\alpha \leq 2t}$, starting from $t=d$. The state $\rho$ is separable if and only if there exists a flat extension $(y_{\alpha})_{\alpha \leq 2(t+d_0)}$ of $(y_{\alpha})_{\alpha \leq 2t}$ with $M_t(y)\geq 0 $ and $ M_t(g_{j}\star y)\geq 0  $ for $ j=1,...,m $.

\subsection{Tms for quantum channels}
\label{thm}

We will now reformulate the theorem above to give a necessary and sufficient criterion for the separability of quantum channels. 
Let $\Phi: \cL(H_{A}\otimes H_{B})\rightarrow \cL(H_{A}\otimes H_{B})$ be a completely positive map and $ C_{\Phi}$ its corresponding Choi matrix acting on $\calH= H_{A}\otimes H_{B}\otimes H_{A'}\otimes H_{B'}  $; an orthogonal basis of $\calH$ is then given by matrices $ S_{\mu_{A}\mu_{B}\mu_{A'}\mu_{B'}}=S^{(A)}_{\mu_{A}} \otimes S^{(B)}_{\mu_{B}} \otimes S^{(A')}_{\mu_{A'}} \otimes S^{(B')}_{\mu_{B'}}$, where $S^{(\bullet)}_{\mu}$ are Hermitian matrices forming an orthogonal basis of the set of bounded linear operators on $ H_{\bullet}$.
 Let us translate the above tms theorems as necessary and sufficient conditions on the Choi matrix to be separable. 
 
 The compact $K$ is defined according to the decomposition we are interested in. In the EB case, one wants to decompose the Choi matrix as $\sum_k P_k\otimes Q_k$, where  $P_k$ and $Q_k$ are positive operators acting on $H_{A}\otimes H_{B}$ and $H_{A'}\otimes H_{B'}  $, respectively. 
Expanding the $P_k$ over a basis of operators $S^{AB}_\lambda$ (these $S^{AB}_\lambda$ could be taken as the $S^{(A)}_{\mu_A} \otimes S^{(B)}_{\mu_B} $) and $Q_k$ over a basis $S^{A'B'}_{\lambda'}$
and expressing the condition that they must be positive, we obtain a definition of the compact $K$ as the set of real
expansion 
coefficients $c_{\lambda},d_{\lambda'}$ such that 
\begin{align}
\label{ineqcd1}
&\sum_{\lambda} c_{\lambda}S^{AB}_\lambda\geq 0,\\
\label{ineqcd2}
&\sum_{\lambda'} d_{\lambda'}S^{A'B'}_{\lambda'}\geq 0.
\end{align}
These positivity conditions can be rewritten as inequalities on the coefficients of the corresponding characteristic polynomials using the Descartes sign rule (see Section \ref{sec2} below). 
In the SEP case, the Choi matrix now has to be decomposed as $\sum_k P_k\otimes Q_k$ with $P_k$ and $Q_k$ acting on $H_{A}\otimes H_{A'}$ and $H_{B}\otimes H_{B'}$, respectively.
The same reasoning applies for the positivity conditions as in the EB case.\\

Given a channel $ \Phi$, we expand the corresponding Choi matrix as 
\begin{itemize}
    \item for EB, 
$C_{\Phi}=\sum_{\lambda,\lambda'}X_{\lambda\lambda'} S^{AB}_{\lambda}\otimes S^{A'B'}_{\lambda'}$ (with $S^{AB}_{\lambda}$ a basis of operators for the system and $S^{A'B'}_{\lambda'}$ for the ancilla) 
\item for SEP, 
$C_{\Phi}=\sum_{\lambda,\lambda'}\tilde{X}_{\lambda\lambda'} S^{AA'}_{\lambda}\otimes S^{BB'}_{\lambda'}$ (with $S^{AA'}_{\lambda}$ a basis of operators for the Hilbert space $H_{A}\otimes H_{A'}  $,  and $S^{BB'}_{\lambda'}$ for the Hilbert space $H_{B}\otimes H_{B'}  $).
\end{itemize}
We can then map either the coordinates $X_{\lambda\lambda'}$ or the coordinates $\tilde{X}_{\lambda\lambda'}$ to a tms $(y_{\alpha})_{\alpha \leq 2}$ (indeed, since we look for separability across a bipartition, the degree of the tms is 2).
The necessary and sufficient conditions for channels are then given as follows:\\\\
{\em Theorem 4}

 (i) The channel $\Phi$ is EB if and only if, considering extensions $(y_{\beta})_{\beta \leq 2t} $ of $(y_{\beta})_{\beta \leq 2} $, there exists a flat extension $(y_{\beta})_{\beta \leq 2(t+d_0)} $ of $(y_{\beta})_{\beta \leq 2t}$ (possibly with $t=1$), with $ M_t(y)\geq 0 $ and $ M_t(g_j\star y)\geq 0  $ for $ j=1,...,m $, where  the $g_j$ are polynomials of variables $c_{\lambda}$ and $d_{\lambda'}$ defined by the conditions $\sum_{\lambda} c_{\lambda}S^{AB}_\lambda\geq 0$, $\sum_{\lambda'} d_{\lambda'}S^{A'B'}_{\lambda'}\geq 0$, and $ d_0=\max_{1\leq j \leq m}\lbrace 1,\lceil \deg(g_j)/2\rceil\rbrace$.
 
(ii) The channel $\Phi$ is SEP if and only if, considering extensions $(y_{\beta})_{\beta \leq 2t} $ of $(y_{\beta})_{\beta \leq 2} $, there exists a flat extension $(y_{\beta})_{\beta \leq 2(t+d_0)} $ of $(y_{\beta})_{\beta \leq 2t}$ (possibly with $t=1$), with $ M_t(y)\geq 0 $ and $ M_t(g_j\star y)\geq 0  $ for $ j=1,...,m $, where  the $g_j$ are polynomials of variables $c_{\lambda}$ and $d_{\lambda'}$ defined by the conditions $\sum_{\lambda} c_{\lambda}S^{AA'}_\lambda\geq 0$, $\sum_{\lambda'} d_{\lambda'}S^{BB'}_{\lambda'}\geq 0$, and $ d_0=\max_{1\leq j \leq m}\lbrace 1,\lceil \deg(g_j)/2\rceil\rbrace$.\\

In the case of fully separable channels, the Choi matrix must be separable across any cut. We expand the matrix $C_{\Phi}$ as $C_{\Phi}=X_{\mu_{A}\mu_{B}\mu_{A'}\mu_{B'}} S^{(A)}_{\mu_{A}} \otimes S^{(B)}_{\mu_{B}} \otimes S^{(A')}_{\mu_{A'}} \otimes S^{(B')}_{\mu_{B'}}$. The coefficients $X_{\mu_{A}\mu_{B}\mu_{A'}\mu_{B'}}$ are now mapped to a tms of order $4$, and the set $ K $ is given by positivity conditions on each Hilbert space. The channel $\Phi$ is fully separable if and only if, looking for extensions of that tms, we find a flat extension (with positivity conditions on the moment and localizing matrices).

\subsection{The algorithm}
\label{sec2}


Theorem 4 can be translated into an algorithm that characterizes separable or entangling channels with respect to a chosen partition.
 The algorithm is based on semidefinite programming (SDP). It takes as only input the corresponding Choi matrix, acting on the system-ancilla Hilbert space $H= H_{A}\otimes H_{B}\otimes H_{A'}\otimes H_{B'}$, whose coordinates (in a basis depending on the partition chosen) provide a tms  $y_\alpha$. The SDP algorithm minimizes a linear function of the moments $y_\alpha$ under the constraints that the moment matrix and the localizing matrices are positive semidefinite.  
 
 In order to define the localizing matrices, the algorithm also requires that the polynomials $g_j$ defining the compact $K$ be specified. They are obtained via positivity conditions for matrices, such as in Eqs.~\eqref{ineqcd1}--\eqref{ineqcd2}. Let $W$ be such a matrix (which depends on the set of variables associated with each Hilbert space, for instance the $c_\lambda$ in Eq.~\eqref{ineqcd1}).  To derive an explicit expression for the $g_j$, we express the coefficients of the characteristic polynomial $p(z)=\sum_{k=0}^n (-1)^{n-k}a_k z^k $ of $W$ through the recursive Faddeev-LeVerrier algorithm, i.e for $1\leq m\leq n$,
\begin{equation}
a_{n-m}=-\frac{1}{m}\sum_{k=1}^m (-1)^k a_{n-m+k}\tr(W^k)
\label{c}
\end{equation}
with $a_n=1$ and $ a_0=\det(W) $. From Descartes sign rule, positivity of $W$ is equivalent to having $a_k\geq 0$ for all $k$.
Let us consider for example the case of 2-qubit channels, for which $ i,j $ go from $ 0 $ to $ 1 $ in  Eq.~\eqref{choisep} and $ C_{\Phi} $ is a $ 16\times 16 $ matrix, and look for its separability as a tensor product of two $4\times 4$ matrices. The characteristic polynomial for each factor is then of degree $ 4 $ ($ n=4 $ in Eq.~\eqref{c}) and the inequalities for positivity  are given by Newton's identities (also known as Girard-Newton formulae). Besides  $a_4=1$ and $a_3=\tr W=1$ (since $W$ is a density matrix), 
we get the conditions 
\begin{align}
& a_2=\frac{1}{2} \left(1- \tr W^2\right)\geq 0 ,\nonumber \\
& a_1=\frac{1}{6} \left(2  \tr W^3 -3  \tr W^2 +1\right)\geq 0 \nonumber \\
& a_0=\frac{1}{24}(-6  \tr W^4 +8  \tr W^3 +3  (\tr W^2)^2-6 \tr W^2 +1)\geq 0,
\label{k}
\end{align}
which yield polynomial inequalities on the $c_\lambda$.

The tms $ y_{\alpha} $ associated with $ C_{\Phi} $ is obtained from its coordinates in a certain basis. In the case of states (see Sec.~\ref{tmsStates}), specifying the coordinates of the density matrix was equivalent to fixing some moments of the measure $d\mu(x)$ as being the expectation values of some physical observables, given by $ \tr(\rho S^{(1)}_{\mu_{1}}\otimes ...\otimes S^{(\nbpart)}_{\mu_{\nbpart}}) $. In the case of channels instead, the observables are relative to the enlarged space system-ancilla, so in order to perform physical measurements on the system only one needs to express the values $ \tr( C_{\Phi} S^{(A)}_{\mu_{A}} \otimes S^{(B)}_{\mu_{B}} \otimes S^{(A')}_{\mu_{A'}} \otimes S^{(B')}_{\mu_{B'}} ) $ in terms of the entries of the superoperator $M$ specifying the channel as  $\rho'_{ij}=M_{ij,kl}\rho_{kl}$. This gives a direct relation with the input-output representation, i.e.~the quantum channel $ \Phi $ is seen as a dynamical process: if $ \rho $ is the initial (input) state before the process, then $ \Phi(\rho) $ is the final (output) state after the process occurs. We can go from one representation to the other considering that $ M $ and $ C_{\Phi} $ are related by the reshuffling operation in the computational basis; for a generic basis this will in general result in a linear combination of physical measurements on the system. The number of physical measurements needed to fix one entry of the moment matrix relative to $ C_{\Phi} $ can be used for instance as a cost function to decide between efficiency of entanglement detection and experimental convenience. The system-ancilla approach is what is used in the so-called Ancilla-Assisted Process Tomography (AAPT) (see e.g. \cite{aapt}), while the input-output one is the Standard Quantum Process Tomography (SQPT) (see e.g. \cite{sqpt}). 

The SDP algorithm then consists in minimizing a function $\sum_\alpha R_\alpha y_\alpha$, with  $R_\alpha$ an arbitrary polynomial, under the constraint that $M_{t}(y)$ and the localizing matrices $M_t(g_j\star y)$ are positive semidefinite, and look for an extension such that the flatness condition is fulfilled. The algorithm is implemented using GloptiPoly \cite{HenLasLoe09} and the MOSEK optimization toolbox \cite{mosek}.
Note that if the rank condition is not met the SDP can still yield a solution to the minimization problem \cite{HenLas05}, but it doesn't tell us anything a priori on the representing measure problem.
To describe all the ingredients in the algorithm, to study its complexity and its efficiency, we will apply it in the next Section to different examples: the spin-$ 1 $ channels mentioned already above, and specific $2$-qubit channels, which are relevant in many experimental settings.

\section{Examples}
In the general case, the number of moments involved, and thus the size of the moment matrices, scales very fast with the extension order $t$, so that numerically the SDP soon becomes intractable. More specifically, while 
full separability of 2-qubit channels 
is a 
problem that is still tractable numerically, already the SEP and EB cases turn out to be too complex if we consider arbitrary qubit channels. Indeed, in that case the variables involved are $(x_\mu)_{1\leq \mu \leq 15} $ for the system and $(x'_\mu)_{1\leq \mu \leq 15} $ for the ancilla. The number of decision variables in the SDP is the number of free entries of the extension of the moment matrix we are looking for; in the order-$t$ extension $M_t(y)$, it is the number of monomials from $30$ variables up to degree $2t$, given by $\binom{30+2t}{2t}$ (see Eq.~\eqref{nrnuples}). Moreover, the polynomials defining the compact $K$ for a two-qubit Hilbert space (of dimension 4) are the ones given at Eq.~\eqref{k}, that is, their degree is 4, and thus $d_0=2$. Since the smallest moment matrix containing all given moments is $M_1(y)$, the smallest extension we have to consider in Theorem 4 is $M_3(y)$. The size of this matrix is $\binom{33}{3}=5456$, and the number of decision variables is $\binom{36}{6}\geq 10^6$. Therefore, the size of the SDP grows very quickly, and thus the number of semidefinite constraints requires too much time and memory. 

Nevertheless, the algorithm can still be applied to families of channels for which the number of variables involved is smaller than in the general case.
In the following we present different examples of such families. We highlight their complexities and computational cost, and explain in more detail the role of the different factors mentioned above. We finally outline some numerical results on their entangling or separable properties.

\subsection{Fully symmetric Choi matrix}
\label{fulsym}
We start with a simple example which allows us to highlight the connection between the TMS algorithm for channels and for states. We  consider quantum channels $\Phi$ such that the Choi matrix $C_{\Phi}$ has components only on the symmetric subspace. In other words, we impose that the 4-qubit state associated with the two-qubit channel $\Phi$ via the Choi-Jamio\l{}kowski isomorphism be fully symmetric under permutation of the qubits  (in the sense that it is a mixture of fully symmetric pure states). In that case, the Choi matrix only has components on the subspace spanned by Dicke states $\ket{D_j^{(m)}}$, which are the symmetrized tensor products of $2j$ qubits, with $j=2$ (4 qubits) and $-j\leq m \leq j$. This means that 
\begin{equation}
(1\!\!1-P) C_{\Phi} (1\!\!1-P) = (1\!\!1-P) C_{\Phi} P = P C_{\Phi} (1\!\!1-P) = 0,
\label{fs}
\end{equation}
where $P=\sum _{m=-2}^2  \vert D_4^{(m)}\rangle\langle D_4^{(m)}\vert $ is the projection operator onto the symmetric subspace.  The constraints in Eq.~\eqref{fs} fix conditions on the superoperator $M$  of which $C_{\Phi}$ is a reshuffling. For $j=2$, only $(2j+1)^{2}$ real independent parameters remain. 

Such a restriction has a clear physical interpretation in the case of one-qubit channels. Indeed, the Choi matrix of a non-unital one-qubit channel can be put in the form 
\begin{equation} \label{ChoiT}
\frac{1}{2}\left(
\begin{array}{cccc}
1+\lambda_3+t_3&0&t_1+it_2&\lambda_1+\lambda_2\\
0&1-\lambda_3+t_3&\lambda_1-\lambda_2&t_1+it_2\\
t_1-it_2&\lambda_1-\lambda_2&1-\lambda_3-t_3&0\\
 \lambda_1+\lambda_2&t_1-it_2&0&1+\lambda_3-t_3 
\end{array}
\right)\,.
\end{equation}
in the canonical basis  \cite{Zycbook}. Imposing that the matrix is associated with a symmetric state is equivalent to imposing that it has no component over the singlet state; this leads to the conditions $t_1=t_2=t_3=0$ (i.e.~the channel is unital) and $\lambda_1-\lambda_2+\lambda_3=1$, which correspond to a face of the tetrahedron of admissible values of the $\lambda_i$ corresponding to unital channels, given by the Fujiwara-Algoet conditions $1\pm\lambda_3\geq |\lambda_1\pm\lambda_2|$ \cite{FujAlg99}. Such points on a face of the tetrahedron correspond to channels whose Kraus rank is 3, which are characterized by the fact that they are the only indivisible channels (that is, they cannot be written as the composition of two non-unitary channels) \cite{Fuji, WolCir08}.


 In the two-qubit channel case there is no such clear geometrical picture of the fully symmetric Choi matrix. However, since the Choi state is a fully symmetric state of $ N=4 $ qubits, if it is separable with respect to an arbitrary partition, then it is fully separable, and it can be written as a convex sum of $N$ projectors on pure symmetric states (see e.g.~\cite{fullysep2}).  
This means that in this case we only need to consider the fully separable case, which coincides with exploring the case of spin-$ 2 $ states (since those states can be seen as symmetric states of 4 qubits).
The tms algorithm for states was exploited in \cite{us} to investigate multipartite entanglement of such states. The problem can be formulated as in Eq.~\eqref{ktms}, with a tms of degree $ 4 $ (thus the smallest moment matrix to consider in Theorem 4 is $M_2(y)$) and a vector of variables $(x_1 , x_2,x_3 )$ (as explained in Section \ref{tmsStates}, since the state is fully symmetric we only need the $ 3 $ variables associated with a single qubit). The semialgebraic set $ K $ is the Bloch sphere, so that $ d_0=1 $. Thus, the first flatness condition in Theorem 4 reads $ \rk M_3(y)=\rk M_{2}(y) $, with $ M_2(y) $ and $ M_3(y) $ of size respectively $ 10\times 10 $ and $ 20\times 20 $. The algorithm usually stops at the first extension and it takes at about $ 1 s $ to give a certificate of separability or entanglement of the channel. We refer to the results obtained for states in \cite{tms17} and \cite{us} for more detail on the implementation in that case.

\subsection{2-qubit planar channels}
\label{planar}


We now consider the case where the $ 2 $-qubit channel is a linear combination of tensor products of single-qubit planar channels. Such one-qubit channels $\phi_{\pl}$ send the (three-dimensional) Bloch ball into a (two-dimensional) ellipse. Note that, according to the so-called "No-Pancake theorem" a planar channel cannot map the Bloch ball to a disk touching the sphere, unless it reduces to a point or a line (see \cite{qeb} and \cite{Fuji}). 

Any one-qubit channel can be described by a $4\times 4$ matrix of the form
\begin{equation}
M=\left(
\begin{array}{cccc}
 1 & 0 & 0 & 0 \\
 t_1 & \text{$\lambda_1 $} & 0 & 0 \\
 t_2 & 0 & \text{$\lambda_2 $} & 0 \\
 t_3 & 0 & 0 & \text{$\lambda_3 $} \\
\end{array}
\right),
\end{equation}
where $ \boldsymbol{\lambda}=(\lambda_1,\lambda_2,\lambda_3) $, with $\lambda_i\geq 0$, is the distortion vector and $\boldsymbol{t}= (t_1,t_2,t_3) $ is the translation vector. Geometrically, the channel maps the Bloch vector $\boldsymbol{r}$ to $M\boldsymbol{r}+\boldsymbol{t}$, that is, the sphere becomes an ellipsoid whose half-axes are given by the $\lambda_i$ and centered at $\boldsymbol{t}$. 

Planar channels are those where one of the $\lambda_i$ is zero.  In \cite{EA2} this type of channels was investigated, but with focus on their entanglement-annihilating properties. In what follows, we consider planar channels $\phi_{\pl}$  with $\lambda_2=0 $ and $ t_2=0 $. 
The condition of complete positivity in the case of a unital planar channel ($\boldsymbol{t}=0$) is given by $\vert \lambda_1 \vert \leq 1- \vert \lambda_3 \vert$, with $\vert \lambda_1 \vert$, $\vert \lambda_3 \vert$ the half-axes of the ellipse;  in the case of non-unital channels the conditions for complete positivity can be found in \cite{Fuji}.

 
Here we investigate whether linear combinations such as
\begin{equation}
    \Phi=a\phi_{\pl}^{(1)}\otimes\phi_{\pl}^{(1)}+b\phi_{\pl}^{(2)}\otimes\phi_{\pl}^{(2)}
    \label{plan}
\end{equation}  
with $ a,b \in \mathbb{R} $ result in separable channels.
We considered the case in which both $\phi_{\pl}^{(1)}$ and $\phi_{\pl}^{(2)}$ are unital, one unital and the other non-unital, and both non-unital. Note that the states \eqref{plan} are not symmetric states in general, as they are symmetrizations of mixed states but not mixtures of symmetric pure states.

The Choi matrix $ C_{\Phi} $, properly normalized ($b=\frac{1}{16}-a$), gives the Choi state on which we apply our algorithm; the basis over which $ C_{\Phi} $ is expanded is chosen as the tensor product $ \sigma_{\mu_1}\otimes \sigma_{\mu_2}\otimes\sigma_{\mu_3}\otimes\sigma_{\mu_4}$ with $ 0\leq \mu_i \leq 2$ and $\lbrace \sigma_{\mu_i}\rbrace=\lbrace 1\!\!1, \sigma_x, \sigma_z \rbrace $, $\sigma_x, \sigma_z$ being the usual Pauli matrices (this is also reasonable from the experimental point of view, since Pauli physical measurements are often used for multi-qubit channels). The Choi states associated with states \eqref{plan} turn out to be equal to their partial transpose with respect to any qubit.
Invariance under partial transposition with respect to the first qubit in $ 2\times N $ systems was shown in \cite{seppt} to entail separability. Therefore the 4-qubit Choi state is separable across any bipartition into sets of 1 and 3 qubits.

Separability for the bipartitions into two sets of 2 qubits, required from the definition of EB and SEP channels, corresponds to the situation of Theorem 4 and can be explored with our algorithm as follows. In contrast to the symmetric case addressed in Subsection \ref{fulsym}, there are now different variables $x_i$ in Eq.~\eqref{def} for the system $A$ and the ancilla $A'$ (and equivalently for $B$ and $B'$)

Let us first consider the question of full separability. In that case, since each system qubit and ancilla qubit is respectively described by two variables $ (x_\mu^A)_{1\leq \mu \leq 2},(x_\mu^B)_{1\leq \mu \leq 2} $ and $ (x_\mu^{A'})_{1\leq \mu \leq 2},(x_\mu^{B'})_{1\leq \mu \leq 2} $, the vector of variables has length $8$. The moments $y_\alpha$ are given by entries of the Choi matrix, the tms has degree $4$, so that formula \eqref{nrnuples} applies with $n=8$ and $2d=4$.
The semialgebraic set is given by the choice of basis matrices for the Choi matrix. Since we expanded it over Pauli matrices, the constraint for each set of variable is the one for qubits, i.e.~the vector of variables is restricted to the Bloch ball. The compact $K$ is therefore the product of 4 unit disks.

Since all polynomials defining $K$ are of degree 2, we have $ d_0=1 $, and thus the first rank condition reads $ \rk M_3(y)=\rk M_{2}(y) $, where the moment matrices have size $\binom{n+t}{t}$, i.e.~respectively $ 165 $ and $ 45 $.
A first hint on the computational complexity of the SDPs 
we need to solve is given by the number of decision variables of the optimization, which in our case corresponds to the number of monomials from $ 8 $ variables up to degree $ 6 $, the latter being the degree of the extension of the tms needed to
construct $M_3(y)$. 
Moreover, SDP are usually solved with the Interior Point Method; each iteration in the primal-dual interior point algorithm requires the solution of a linear system, which is the most expensive operation with $ O(N^3) $ complexity, solvable using Gaussian elimination. Here $ N $ is the number of linear constraints in the SDP and efficiency drops with the growing number of semidefinite terms involved in these linear constraints, which in the case here considered are $ \sim 10^3 $. This in general has a big impact on the time and  memory requested for a single run of the algorithm \cite{mosek}. 
Nevertheless, we could run our algorithm in that case, which allowed us to test for separability of channels of the form \eqref{plan}. The algorithm still performs very well; for all the examples tested a certificate of separability was found either at the first relaxation order $\rk M_3=\rk M_2$ (with a time of $\sim 10 s$ for a single run) or at the second relaxation order $\rk M_4=\rk M_3$ (with a running time of $\sim 6$min).

All the Choi states tested result fully separable for all the three cases listed above (where channels $\phi_{\pl_i}$ can be unital or not); as a consequence, all these states are both EB and SEP. Based on the available numerical evidence, we conjecture that all states of the form (20) are fully separable.

\subsection{Qutrit channels}

We now study the case of qutrit channels. More specifically, we apply our algorithm to a family of channels presented in \cite{Che}, where EB properties of qutrit gates were studied through the negativity $ N(\rho)=\frac{1}{2}(\Vert \rho^{T_H}\Vert_1-1) $, with $ \Vert \rho^{T_H}\Vert_1 $ the trace norm of the partial transpose with respect to the system qutrit. The negativity $ N(\rho) $ cannot detect PPT-entangled states; in other words there exist entangled states with $ N(\rho)=0 $. For such states, our algorithm is able to give a certificate of separability, as we illustrate below. Note that, even though in this case the system is not bipartite, the definition of entanglement breaking still applies since it involves the presence of an ancilla, as explored for 1-qubit channels in \cite{qeb}; on the other hand, the definition of SEP separability cannot be applied to this example.


As a basis for qutrit density operators, we use Gell–Mann matrices $ \lbrace \lambda_i\rbrace_{i=1}^8 $ together with  $ \lambda_0=\sqrt{\frac{2}{3}}\mathbb{1} $. In this basis, an arbitrary qutrit density matrix can be written as
\begin{equation}
\rho=\frac{1}{3}(\mathbb{1}+\sum_{i=1}^8 \zeta_i \lambda_i) 
\label{rhoqutr}
\end{equation} 
with $ \zeta_i=\frac{3}{2}\tr(\rho \lambda_i) $.

\begin{figure}
\includegraphics[scale=0.7]{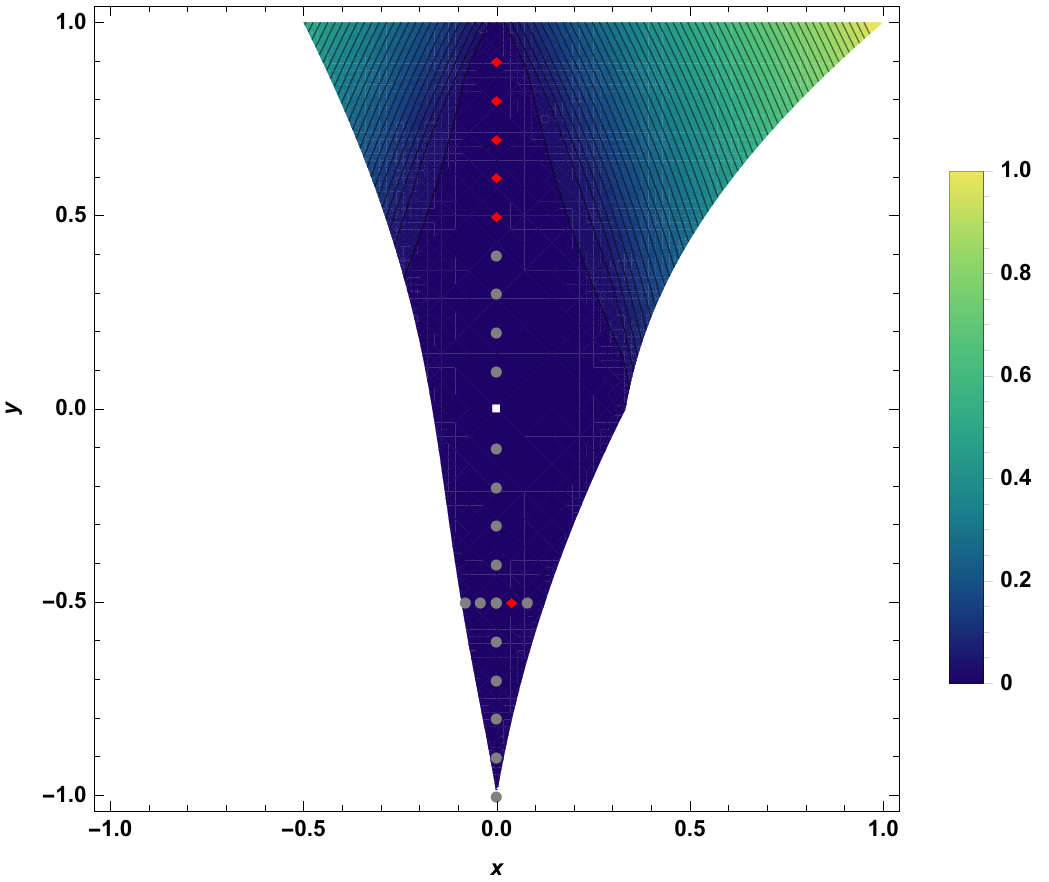}
\caption{Region of $ x $ and $ y $ parameters for which $ C_{\Phi_D} $
of the damping qutrit channel 
is positive semidefinite; the color function corresponds to the negativity values in the range $ [0,1] $, with steps for the contour lines of $0.02$. The central plateau corresponds to the region of zero negativity, where the PPT criterion remains inconclusive. Grey points correspond to states found separable by our algorithm,
signifying entanglement breaking channels;
red points correspond to states where the algorithm needs to go to a higher extension order
and remains inconclusive with our numerical resources. 
The white point 
marks  the maximally mixed state.}
\label{fig2}
\end{figure}

The channel we consider is a damping qutrit channel, i.e. a channel that can be written as an affine
transformation on the generalized (qutrit) Bloch vector as $ \Phi_D: \boldsymbol{\zeta}\rightarrow \boldsymbol{\zeta}'=\Lambda \boldsymbol{\zeta}$, where $ \Lambda=\textrm{diag}(\Lambda_1,...,\Lambda_8) $ is the damping matrix. The $ \Lambda_i $ cannot take any arbitrary value because $ \Phi_D $ has to be completely positive, thus leading to the constraints $ \vert \Lambda_i\vert \leq 1 $. More specifically, we consider the family of damping channels given in \cite{Che} and parametrized by $ \Lambda_{i\neq 3,8}=x, \Lambda_{i=3}=y,\Lambda_{i=8}=y^2 $. The Choi state corresponding to $ \Phi_D $ can be written by transforming the propagator to the canonical basis, then reshuffling and normalizing (it corresponds to a maximally mixed state for $ x=y=0 $ and to a maximally entangled state of two qutrits for $ x=y=1 $). The region of parameters for which $ C_{\Phi_D} $ is positive semidefinite together with the values of the corresponding negativity is shown in Fig.~\ref{fig2}.

Any two-qutrit state can be expanded over the basis formed by tensor products of Gell-Mann matrices \cite{qutrstates}. This setting is analogous to the one described in section \ref{tmsStates} for two spin-$1$ states. The vector of variables is $x=(x_1,x_2,...,x_{16})$, where $x_1,...,x_8$ are the coordinates $\alpha_i$ associated with the system qutrit, and $x_9,...,x_{16}$ are associated with the ancilla qutrit. Since there are two subsystems, the tms has degree $2$. The characteristic polynomial for a qutrit density matrix has degree 3, therefore the semialgebraic set is given by the conditions $\tr\rho^2\leq 1$ and $\det\rho\geq 0$, with $\rho$ the density operator in Eq.~\eqref{rhoqutr}. It follows that the corresponding polynomials of the variable $x_i$ have maximal degree 3, and thus $d_0=2 $. This gives the rank shift in Theorem 4: at the first iteration of the algorithm the flatness condition reads $ \rk M_3(y)=\rk M_1(y)$. These moment matrices have size respectively $ 969 $ and $17$. The number of decision variables in the SDP corresponds to the number of monomials from $ 16 $ variables up to degree $ 6 $ ($ \sim 7\times 10^4 $) and the number of semidefinite constraints is given by $ \binom{n+t}{t}+m\binom{n+t-1}{t-1}+m\binom{n+t-2}{t-2} $,
that is, the size of the moment matrix of the first extension ($ t=3 $) and the size of the localizing matrices multiplied by the number $ m $ of inequalities in the semialgebraic set for each set of variables.

 The tms algorithm can be exploited to investigate in particular the Choi states with zero negativity, for which the PPT criterion alone is inconclusive. The results for some pairs of parameters with $ (x=0,y\in [-1,1]) $ and $ (x\in [-\frac{2}{25},\frac{2}{25}],y=-\frac{1}{2}) $ are explored and they are shown in Fig.~\ref{fig2}. The points highlighted in grey are the points tested with the algorithm which give a certificate of separability, including the white point which corresponds to a Choi state equal to the maximally mixed state of two qutrits. In the latter cases the SDP is feasible and the flatness condition $ \rk M_3(y)=\rk M_1(y)$ is satisfied, meaning that the corresponding $ \Phi_D $ are EB; on the other hand, the algorithm remains inconclusive for the red points at the first iteration, leading to the necessity for higher-order extensions, which are 
 beyond our computational resources. We did not detect PPT entangled states among the tests done; the algorithm confirms entanglement for negativity greater than zero for all the states tested.
A single run of the algorithm in this case takes about $5h$ and between $150$ and $300$ GB of RAM. 


 

\section{Conclusions}
In this paper we have discussed an algorithm that deterministically detects whether a quantum channel is separable or not, or whether it is entanglement breaking or not. We explored three classes of separability across different cuts between systems and ancillae (SEP, EB or FS). This algorithm is based on a mapping between coordinates of the Choi matrix of the channel, expressed in a given basis, and a truncated moment sequence. Low-order moments are fixed by measurements performed on the channel, and the separability problem is equivalent to finding whether these moments are those of a measure supported on a certain compact set.

In the case of fully symmetric 
Choi matrices for qubit channels, 
where the aim is to find a decomposition over the Bloch sphere, the number of variables in the tms is $n=3$, so that the size of a moment matrix of order $t$ is $\binom{n+t}{t}\sim t^3/6$. On the other hand, in the simplest case of detection of EB or SEP in a generic two-qubit channel, there are $n=30$ variables involved, and thus the size of the moment matrix is $\binom{n+t}{t}=5456$ for $t=3$. Moreover, the number of independent entries in $M_t(y)$ is given by $\binom{n+2t}{2t}\sim 2\times 10^6$ for $t=3$.
Nevertheless, we can consider families of channels for which the number of free parameters in each subsystem is smaller than in the general case. Then, the number of variables involved in the mapping to tms is reduced and the 
matrices in the SDP become amenable to numerical investigation.
As we showed here, this is the case for planar channels (where one dimension is suppressed) or qutrit channels (which live in the symmetric space of two qubits). Our algorithm is then able to decide whether the channel is EB or SEP. For instance in the case of qutrit channels we were able to provide a certificate of separability in cases where the negativity of the Choi matrix 
vanishes and thus 
is unable to yield a conclusion. Since calculations are costly, this approach could be used as a numerical tool to explore possible conjectures or produce counter-examples.

\appendix

\section{Sketch of the proof of Theorem 3}
\label{proofth3}
Suppose $\rk M_t(y)=r$ with $M_t(y)\geq 0$ and there exists a flat extension $M_{t+d_0}(y)$ with $M_t(g_j\star y)\geq 0$ for $1\leq j \leq m$. Then $M_{t+1}(y)$ is also a flat extension of $M_t(y)$, and we then know from Theorem 2 that $y$ admits a (unique) $r$-atomic representing measure supported by $\bx_k\in \mathcal{V}(\ker M_t(y))$.
All what remains to show is that positivity of the localizing matrices enforces that the $\bx_j$ belong to $K$, that is, $g_j(\bx_k)\geq 0$ for $1\leq j \leq m$ and  $1\leq k \leq r$. 

This can be done as follows. First, observe that since $M_t(y)$ is of rank $r$, one can find a nonsingular $r\times r$ principal submatrix of $M_t(y)$. If $\mathcal{B}$ is the set of labels $\alpha$ of the  
rows of that matrix, then the image of  $M_t(y)$ is spanned by the $\bx^\alpha$, $\alpha\in\mathcal{B}$, and by definition these $\bx^\alpha$ are of order less than or equal to $t$. Since the whole vector space of polynomials can be decomposed as a direct sum of the image and the kernel of $M_t(y)$, an arbitrary polynomial $p$ can be decomposed as $p=q+\tilde{p}$ with $q=\sum_{\alpha\in\mathcal{B}}q_\alpha\bx^\alpha\in\im M_t(y)$ and $\tilde{p}\in\ker M_t(y)$. 

Now let  $p_k$ be interpolating polynomials of the $\bx_{k'}$, which are the atoms supporting the representing measure of $y$. That is, 
 $p_k(\bx_{k'})=\delta_{kk'}$ for $1\leq k,k' \leq r$. One can decompose them as above as $p_k=q_k+\tilde{p_k}$ with $\tilde{p_k}\in \ker M_t(y)$ and $q_k$ of degree less than $t$. 
 By definition, the $\bx_{k'}$ are roots of all polynomials in $\ker M_t(y)$, and thus one has $\tilde{p_k}(\bx_{k'})=0$, which implies
 $q_k(\bx_{k'})=\delta_{kk'}$ for $1\leq k,k' \leq r$.

Now, for $y=\int x^\alpha d\mu(x)$ and for arbitrary polynomials represented by vectors $p,q\in \mathbb{R}^{S_{t}}$
\begin{eqnarray}
q^T M_t(y) p&=&q_\alpha M_{\alpha\beta}p_\beta\nonumber\\
&=&q_\alpha y_{\alpha+\beta}p_\beta\nonumber\\
&=&\int q_\alpha x^{\alpha+\beta}p_\beta d\mu(x)\nonumber\\
&=&\int p(x)q(x)d\mu(x)
\end{eqnarray}
(with Einstein summation convention) and
\begin{eqnarray}
q^T M_t(g*y) p&=&q_\alpha g_\gamma y_{\alpha+\beta+\gamma}p_\beta\nonumber\\
&=&\int q_\alpha g_\gamma p_\beta x^{\alpha+\beta+\gamma} d\mu(x)\nonumber\\
&=&\int p(x)q(x)g(x)d\mu(x).
\end{eqnarray}
Thus, $M_t(g_j\star y)\geq 0$ and $d\mu(\bx)=\sum_i \omega_i \delta(\bx-\bx_i)d\bx$
entails $\forall k,j$
\begin{eqnarray}
0\leq  q_k^T M_t(g_j\star y)q_k&=&\int q_k(\bx)^2g_j(\bx)d\mu(\bx)\nonumber\\
&=&\sum_{i=1}^r \omega_i \int d\bx q_k(x)^2g_j(x)\delta(\bx-\bx_i)\nonumber\\
&=&\sum_{i=1}^r \omega_i q_k(\bx_i)^2 g_j(\bx_i)\nonumber\\
&=&\omega_k g_j(\bx_k)\nonumber\\
\end{eqnarray}
since $q_k(\bx_i)=\delta_{ki}$. As all $\omega_k>0$ this implies that $g_j(\bx_k)\geq 0$
and thus $\bx_k\in K$, which completes the proof.\\

\section{Rank property of extensions}
\label{proofrank}
Let us show that the rank condition $\rk M_{t'}(y)=\rk M_{t}(y)$ implies the fact that positivity of  $M_{t}(y)$ and $M_{t'}(y)$ are equivalent. 

Since $M_{t}(y)$ is a principal submatrix of $M_{t'}(y)$ one direction is obvious. To show the converse, suppose $M_t(y)\geq 0$ and $\rk M_t(y)=r=\rk M_{t'}(y)$. Then, as in Appendix \ref{proofth3}, there exists a nonsingular $r\times r$ principal submatrix of $M_t(y)$ indexed by labels $\alpha\in\mathcal{B}$ with $|\alpha|\leq t$. This $r\times r$ submatrix is also a nonsingular principal submatrix of $M_{t'}(y)$. Since $M_{t'}(y)$ has rank $r$, the corresponding $r$ monomials $x^\alpha$ are therefore a basis of $\im M_{t'}(y)$. Since the submatrix is positive because $M_t(y)$ is, then so is $M_{t'}(y)$. 

\bibliographystyle{apsrev}
\bibliography{chanbib}

\end{document}